\documentclass[12pt]{JHEP3}

\usepackage{amssymb,amsmath}
\usepackage{epsfig}

\newcommand{\beq}{\begin{equation}}
\newcommand{\eeq}{\end{equation}}
\newcommand{\baq}{\begin{eqnarray}}
\newcommand{\eaq}{\end{eqnarray}}
\newcommand{\mc}[1]{\mathcal{#1}}

\title{Non-Gaussianity in Curvaton Models with Nearly Quadratic Potential}

\author{Kari Enqvist$^{1,2,}$\footnote{E-mail: kari.enqvist@helsinki.fi} ~and Sami Nurmi$^{2,}$\footnote{E-mail: sami.nurmi@helsinki.fi}\\
${}^1$ Helsinki Institute of Physics, P.O. Box 64, FIN-00014 University of Helsinki, Finland\\
${}^2$ Department of Physical Sciences, P.O. Box 64, FIN-00014
University of Helsinki, Finland}

\abstract{We consider curvaton models with potentials that depart
slightly from the quadratic form. We show that although such a small
departure does not modify significantly the Gaussian part of the
curvature perturbation, it can have a pronounced effect on the level
of non-Gaussianity. We find that unlike in the quadratic case, the
limit of small non-Gaussianity, $|f_{NL}|\ll1$ , is quite possible
even with small curvaton energy density $r\ll1$ . Furthermore,
non-Gaussianity does not imply any strict bounds on $r$ but the
bounds depend on the assumptions about the higher order terms in the
curvaton potential.}

\preprint{HIP-2005-33/TH}


\begin{document}


\section{Introduction}

Possible non-Gaussian features of the Cosmic Microwave Background
(CMB) temperature anisotropy can provide important constraints on
models of inflation. For instance, an observation of a significant
non-Gaussianity would effectively rule out single-field inflation
(for a review of non-Gaussianity, see \cite{review}). Usually in the
literature the non-Gaussianities are characterized by a
non-linearity parameter $f_{NL}$, which is a measure of the
non-Gaussian curvature perturbation relative to the Gaussian
perturbation. Present WMAP observations yield the limit
\cite{WMAP-nongaussian} $-58<f_{NL}<134$ at $95\%$ confidence level.
With polarization measurements, the Planck Surveyor Mission is
expected to push the limit down to  $|f_{NL}|\lesssim 2.9$
\cite{limits}. For the single field inflation one obtains $f_{NL}$
which is of the order of the slow-roll pararameters
\cite{maldacena}; hence a detection of non-Gaussianity by Planck
would indeed suffice to rule out single-field inflation.

For the curvaton models \cite{sloth,moroi,generating} it has been
suggested \cite{curvatonnonG} that a non-observation of
non-Gaussianity would indicate that the model is ruled out, at least
in the case of the quadratic potential. In curvaton models the
curvature perturbation is generated after inflation by the decay of
an effectively massless scalar field $\sigma$ different from the
inflaton. The curvaton energy density remains subdominant until the
end of inflation so that the density parameter\footnote{Our
definition of $r$ is adopted from \cite{generating,bartolo} and
differs from that of e.g \cite{prediction}.} $r\equiv
4\rho_\sigma/(4\rho_r +3\rho_\sigma)\ll 1$, where $\rho_r$ is the
energy density of radiation after inflation. Thereafter the curvaton
field begins to oscillate and behaves effectively like matter. Its
relative energy density grows during oscillations and when it
eventually decays, the perturbation it has received during inflation
will be imprinted on the decay products, the light degrees of
freedom. Because the curvaton is massless, the perturbation is
predominantly Gaussian. However, there will also be a non-Gaussian
contribution that arises because of the curvaton dynamics after
inflation. By now a well-known result is that in curvaton models
with quadratic potential the non-linearity parameter can be written
as \cite{curvatonnonG, on} \beq\label{quadrfnl} f_{NL}=\frac 5 3
+\frac 5 8 r -\frac{5}{3r}~. \eeq Thus, for low $r$ as required for
the subdominance of the curvaton during inflation, $|f_{NL}|$ is
typically much bigger than 1. However, we wish to point out that
this result is considerably modified for non-quadratic potentials.
Although the curvaton must be weakly self-interacting, it is highly
likely that there is some departure from the quadratic potential. As
we will discuss in this paper, while leaving the Gaussian
perturbation essentially unchanged, a small correction to the
quadratic potential can have an important effect on the
non-Gaussianity parameter. Indeed, we show that if one does not
insist on a strictly quadratic potential, it is quite possible to
have $|f_{NL}|$ much less than 1 also in the curvaton scenario.


\section{The curvature perturbation generated in the curvaton model}

 We adopt here the non-linear, the so-called separate universe approach presented
in \cite{prediction,sasaki,wands,proof}, which is valid on large
scales. There one considers perturbations around the homogeneous and
isotropic flat FRW-universe assuming that their spatial variation
outside horizon is smooth. The evolution of large scales is
approximated by replacing each quantity by its spatial average
inside some smoothing scale $k^{-1}_{\rm smooth}$ and considering
these smoothed, locally homogeneous and isotropic regions to evolve
like separate FRW-universes \cite{prediction,wands,proof}. The
spatial variation of perturbations outside the smoothing scale is
taken into account by doing first order gradient expansion leading
to different expansion rates in different smoothed regions. Here we
briefly recapitulate the main features \cite{prediction,proof} of
this approach relating to non-Gaussianity in curvaton models.

In the first order in gradient expansion, the spatial part of the
metric can be written as \cite{proof} \beq\label{metric}
g_{ij}=a^2(t)e^{-2\psi(t,{\bf x})}\gamma_{ij}\eeq where
$\gamma_{ij}$ is constant, assuming the amplitude of the
gravitational waves to be small. The curvature perturbation
$\psi(t,{\bf x})$ thus defined can be interpreted as a perturbation
in the scale factor, $\tilde{a}(t,{\bf x})\equiv a(t)e^{-\psi(t,{\bf
x})}$. As it has been shown in \cite{prediction,proof}, the
curvature perturbation on uniform energy density hypersurfaces stays
constant outside the horizon in the absence of non-adiabatic
pressure perturbation; just like its counterpart in the usual first
\cite{firstorder} and second order perturbation theories
\cite{secondorder}.

The amount of expansion along the worldline of a comoving observer
from a spatially flat $\psi=0$ slice at time $t_1$ to a generic
slice at time $t$ is given by $N(t,{\bf x})={\rm
ln}\frac{\tilde{a}(t,{\bf x})}{a(t_1)}$ since the expansion in a
spatially flat gauge corresponds to that of the unperturbed
universe. By choosing the slice at time $t$ to have uniform energy
density, the curvature perturbation on that slice can be written as
\cite{prediction,proof} \beq\label{zeta}\zeta(t,{\bf
x})\equiv-\psi(t,{\bf x})={\rm ln}\frac{\tilde{a}(t,{\bf
x})}{{a}(t)}=N(t,{\bf x})-N(t) \eeq where $N(t)$ is the amount of
expansion in the background universe.

Following \cite{prediction} we expand the curvature perturbation
(\ref{zeta}) up to second order in the Gaussian curvaton
perturbations in order to take into account the non-Gaussian
effects:\beq\label{expand}\zeta(t,{\bf
x})=\partial_{\sigma}N(t)\delta\sigma(t,{\bf
x})+\frac{1}{2}\partial^2_{\sigma}N(t)\delta\sigma(t,{\bf x})^2.
\eeq In the limit $r\ll1$ the curvature perturbation is almost
completely generated during oscillations of the curvaton field.
Assuming sudden decay at $t=t_{\rm dec}$ the amount of expansion in
the background universe during oscillations is given by
\cite{prediction} \beq\label{N}N(\sigma_{\rm dec},\sigma_{\rm
osc})=\frac{1}{3}{\rm ln}\frac{\frac{1}{2}m_{\sigma}\sigma^2_{\rm
osc}}{\frac{1}{2}m_{\sigma}\sigma^2_{\rm dec}}=\frac{1}{4}{\rm
ln}\frac{(\rho_r)_{\rm osc}}{\rho_{\rm dec}-(\rho_{\sigma})_{\rm
dec}} \eeq where $\sigma_{\rm osc}$ is the value of the curvaton at
the onset of oscillations. The total curvature perturbation is
obtained by substituting Eq. (\ref{N}) into Eq. (\ref{expand})
\baq\label{curvatonzeta}\zeta(t,{\bf x})&=&\frac{r\sigma'_{\rm
osc}}{2\sigma_{\rm
osc}}\delta\sigma_*+\frac{1}{4}\Big(\Big(-\frac{3}{8}r^3-r^2+r\Big)
\Big(\frac{\sigma'_{\rm
osc}}{\sigma_{\rm osc}}\Big)^2+r\frac{\sigma''_{\rm
osc}}{\sigma_{\rm osc}}\Big)\delta\sigma^2_{*} \eaq where we have
denoted $\partial_{\sigma_*}\equiv'$ . The initial values for the
curvaton field $\delta\sigma_*(t,{\bf x})$ in each smoothed region
are set by inflation and the derivatives in Eq. (\ref{expand}) are
thus taken w.r.t the field value during inflation $\sigma_{*}$.

We point out that the nonlinear part in the curvature perturbation,
Eq. (\ref{curvatonzeta}), consists of a square of the linear part
$(\frac{\sigma'_{\rm osc}}{\sigma_{\rm osc}}\delta\sigma_*)^2$ and
of an additional dynamical term. One could thus expect that the
non-Gaussian effects would be more dependent on the dynamics than
the Gaussian part which is indeed quite reasonable since, loosely
speaking, the Gaussian part depends on the size of the perturbations
while the non-Gaussian part depends on the relative size of the
perturbations compared to the background field.

From Eq. (\ref{curvatonzeta}) the curvature perturbation can be
written in the form \cite{komatsu}
$\zeta=\zeta_g-\frac{3}{5}f_{NL}(\zeta_g^2-\langle\zeta_g^2\rangle)$
where $\zeta_g$ is Gaussian and the non-linearity parameter $f_{NL}$
is independent of position. The non-linearity parameter can now be
directly read off from Eq. (\ref{curvatonzeta}):
\beq\label{fnl}f_{NL}=\frac{5}{3}+\frac{5}{8}r-\frac{5}{3r}\Big(1+\frac{\sigma''_{\rm
osc}\sigma_{\rm osc}}{\sigma'^2_{\rm osc}}\Big). \eeq Although this
result was obtained in \cite{prediction,dhlyth}, the dependence of
the last term ${\sigma''_{\rm osc}\sigma_{\rm osc}}/{\sigma'^2_{\rm
osc}}$ on the potential has not been previously examined. In the
quadratic case ${\sigma''_{\rm osc}\sigma_{\rm osc}}/{\sigma'^2_{\rm
osc}}=0$ but, as we will show shortly, this result may be
considerably modified even if the deviation from the quadratic
potential is small.


\section{Small deviation from the quadratic
potential}

The curvaton equation of motion during radiation domination is given
by \beq
\label{feom}\ddot{\sigma}+\frac{3}{2t}\dot{\sigma}+V'(\sigma)=0 \eeq
where we have ignored spatial gradients which are small on large
scales. We now consider the potential of the  form
\beq\label{ourpot} V(\sigma)=\frac{1}{2}m^2_{\sigma}\sigma^2+\lambda
m^{4-n}_{\sigma}\sigma^n. \eeq with $\lambda\ll 1$. To describe the
size of the potential correction at the end of inflation we
introduce a parameter
$s\equiv2\lambda({\sigma_{*}}/{m_{\sigma}})^{n-2}$. The smallness of
the correction in Eq. (\ref{feom}) requires $s\ll {2}/{n}$. In the
quadratic case the e.o.m (\ref{feom}) is nothing but a Bessel
equation with a general solution
$\sigma_0(t)=A_0{J_{(1/4)}(m_{\sigma}t)}/{(m_{\sigma}t)^{1/4}}+
B_0{J_{(-1/4)}(m_{\sigma}t)}/{(m_{\sigma}t)^{1/4}}\equiv
A_0y_1(t)+B_0y_2(t)$. To obtain a regular solution at
$m_{\sigma}t=0$, we must set $B_0=0$. Furthermore, requiring
$\sigma_0(m_{\sigma}t=0)=\sigma_{*}$ we find \beq
\label{sigma0}\sigma_0(t)=\sigma_{*}\frac{\pi}{2^{5/4}\Gamma{(3/4)}}
\frac{J_{(1/4)}(m_{\sigma}t)}{(m_{\sigma}t)^{1/4}} \eeq from which
we see that ${\sigma''_{\rm osc}\sigma_{\rm osc}}/{\sigma'_{\rm
osc}}=0$ for the quadratic case as claimed above.

Let us now make an Ansatz of the form
$\sigma=\sigma_0+\lambda\sigma_1$ in Eq. (\ref{feom}). At first
order in $\lambda$ we obtain the linearized equation of motion \beq
\ddot{\sigma}_1+\frac{3}{2t}\dot{\sigma}_1+m_{\sigma}^2\sigma_1+
m_{\sigma}^{4-n} n\sigma_0^{n-1}=0. \eeq The solution to the
homogeneous equation is already given above; the general solution
$\sigma(t)=A(t)y_1(t)+B(t)y_2(t)$ is obtained by the method of
variation of parameters. The coefficients
 are solved from the equations
\baq
\label{B} A'y_1+B'y_2&=&0\\
\label{A} (A'y_1'+B'y_2')m_{\sigma}^2&=&-m_{\sigma}^{4-n}n
C^{n-1}y_1^{n-1} \eaq where $\frac{d}{d(m_{\sigma}t)}\equiv'$ and
$C\equiv\sigma_{*}\frac{\pi}{2^{5/4}\Gamma{(3/4)}}$. Since the
correction to the quadratic potential is small at the end of
inflation and even smaller at later times we can assume that the
beginning of the oscillations takes place at the same time as in the
purely quadratic case. Thus the era between the end of inflation and
the beginning of oscillations corresponds to $m_{\sigma}t=0\dots 1$.
The equations (\ref{B}), (\ref{A}) can now be solved by expanding
the homogeneous solutions and by a straightforward calculation we
find that up to order $\mc{O}(({m_{\sigma}t}/{2})^8)$
\baq\label{asol}
A(m_{\sigma}t=1)&\approx&-n{\sigma_{*}^{n-1}}{m_{\sigma}^{2-n}}\Big(1.0-0.10n+6.9\times
10^{-3}n^2- \nonumber\\&& 3.6\times 10^{-4}n^3+1.4\times
10^{-5}n^4\Big)\\\label{bsol}
B(m_{\sigma}t=1)&\approx&n{\sigma_{*}^{n-1}}{m_{\sigma}^{2-n}}\Big(0.82-0.095n+6.8\times
10^{-3}n^2 \nonumber\\&& -3.6\times 10^{-4}n^3+1.3\times
10^{-5}n^4\Big)~. \eaq Thus the value of the curvaton at the onset
of oscillations reads \beq\label{curvatonvalue}\sigma_{\rm
osc}\approx0.81\sigma_{*}+{\lambda n m_{\sigma}^{2-n}
}\sigma_{*}^{n-1}g(n) \eeq with $g(n)\equiv-0.20+1.2\times
10^{-2}n-6.1\times 10^{-4}n^2+3.1\times 10^{-5}n^3-2.1\times
10^{-6}n^4$. With the exponent values $n\lesssim10$ we are
considering, $g(n)$ is negative and roughly constant,
$g(n)\sim-0.1$. The non-linearity parameter, valid for any potential
of the type given in Eq. (\ref{ourpot}), is now obtained by
substituting Eq. (\ref{curvatonvalue}) into Eq. (\ref{fnl}):
\baq\label{Fnl}
f_{NL}&=&\frac{5}{3}+\frac{5}{8}r-\frac{5}{3r}\Big(1+n(n-1)(n-2)
\frac{g(n)(0.41s+0.25s^2ng(n))}{(0.81+0.50n(n-1)sg(n))^2}\Big). \eaq


\section{The behaviour of $|f_{NL}|$ in the limit $r\ll1$}

It is readily seen that the effect of the potential correction in
Eq. (\ref{Fnl}) is most significant in the limit of small curvaton
energy density $r\ll1$. In the following we are working in this
limit if not otherwise stated. Thus we can consider the dominating
$1/r$ part of $f_{NL}$ alone and neglect the small contribution from
the remaining terms\footnote{We will show that the $1/r$ term may
vanish in which case other terms in Eq. (\ref{fnl}) should also be
taken into account. The term linear in $r$ is however negligible in
the limit $r\ll1$ and the constant part $5/3$ does not affect our
qualitative conclusions.} $\frac{5}{3}+\frac{5}{8}r$. We now examine
the effect of the potential correction by keeping $s$ fixed. Our
solution to the equation of motion (Eq. \ref{curvatonvalue}) is
constructed in such way that the value of the curvaton during
inflation, $\sigma_{*}$, is fixed to be the same as in the quadratic
case; this is an approximation which however should be justified as
we are considering only small departures from the quadratic
potential. Since we also keep the mass $m_{\sigma}$ unaltered, a
constant $s$ means that we are considering the coupling constant as
a function of the exponent $\lambda=\lambda(n)$.

As stated above, our perturbative approach to solving the equation
of motion (\ref{feom}) puts limits $\lambda\ll1$, $s\ll{2}/{n}$.
Furthermore, the dominance of the Gaussian perturbations and the
masslessness of the curvaton during inflation also restrict the
possible values of $\sigma_{*}$ and $m_{\sigma}$. Using Eq.
(\ref{curvatonvalue}) we find the spectrum related to the two-point
correlator of the curvature perturbation (i.e. Gaussian spectrum) to
be \beq\label{mcp} \mc{P}_{\zeta}=\Big(\frac
{0.81+\frac{sn(n-1)}{2}g(n)}{0.81+\frac{sn}{2}g(n)}\Big)^2\Big(\frac{rH_{*}}{4\pi\sigma_{*}}\Big)^2
\eeq where $\Big({rH_{*}}/{(4\pi\sigma_{*})}\Big)^2$ is the purely
quadratic result; $H_*$ is the Hubble parameter during inflation. In
the small correction limit the prefactor in Eq. (\ref{mcp}) is
$\sim\mc{O}(10^{-1})$ and hence we conclude that, although the
perturbation amplitude is suppressed, the Gaussian part is not
significantly affected by the correction as the suppression can be
compensated by a slight increase in the scale of inflation. Thus we
may use the results for the quadratic parameters \cite{bartolo}
which typically imply the restriction $m_{\sigma}\ll H_{*}\lesssim
\sigma_{*}$ coming from the masslessness of the curvaton field and
the assumed Gaussianity of the curvaton perturbations. This means
that the smallness of the potential correction, $s\ll {2}/{n}$,
requires $\lambda\lll 1$ when $n\gtrsim 4$. Moreover, the
inflaton-generated curvature perturbation is supposed to be
negligible which also requires $\lambda\lll1$ for non-renormalizable
terms, $n>4$.

In Fig. \ref{fnlplot} we show the behaviour of the dominant part of
the non-linearity parameter as a function of $n$ for two selected
values of $s$.
%
%
\EPSFIGURE[!h]{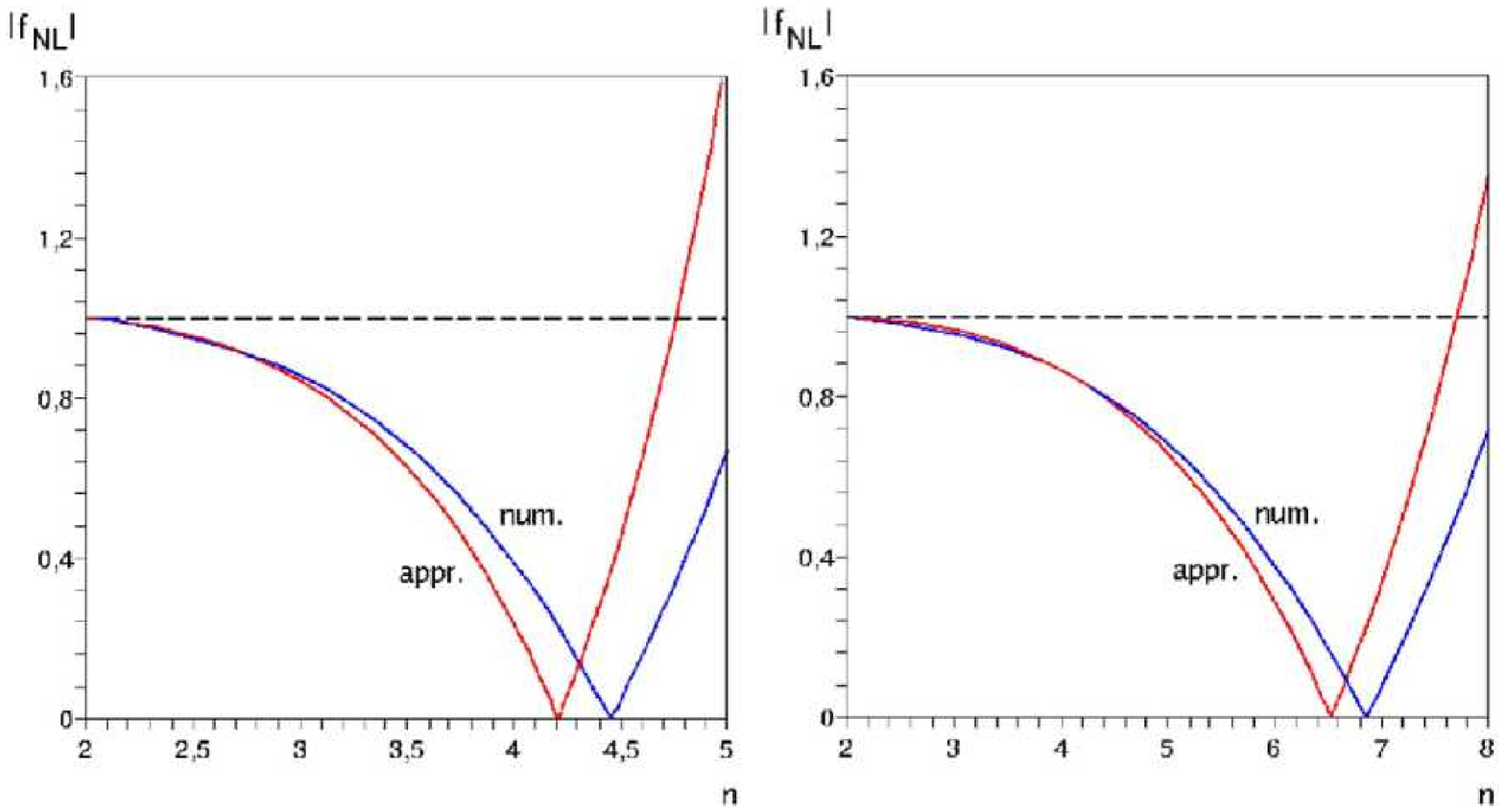,width=11.5cm,height=5.5cm}
{\label{fnlplot}\small The qualitative behaviour of the
non-linearity parameter with $s=0.2$ on the left panel and $s=0.05$
on the right panel; the values of $|f_{NL}|$ are in units $5/3r=1$.
Apart from the analytical result of Eq. (\ref{Fnl}) we also show a
numerical result obtained directly from Eqs. (\ref{fnl}),
(\ref{feom}) setting $t=1/m_{\sigma}$ as the beginning of the
oscillations.}
%
%
%
The value of the non-linearity parameter depends on the parameters
$\lambda,\sigma_{*},m_{\sigma}$, but nevertheless Fig. \ref{fnlplot}
reveals the generic qualitative behaviour of $|f_{NL}|$ as $n$ is
varied. One can clearly see that when the exponent of the potential
correction is increased, the amount of the non-Gaussianity first
begins to decrease as compared to the quadratic case. However, at
large $n$ the value of $|f_{NL}|$ begins to grow rapidly.

The explanation for the behaviour of $|f_{NL}|$ and the physics
involved is most transparent if we switch to the perturbative point
of view. Using perturbation theory one finds \cite{on,primordial}
that the $1/r$ part in Eq. (\ref{fnl}) represents the first order
contribution while the rest is due to second order effects. Thus the
first order theory is adequate as long as we restrict ourselves in
the region where the $1/r$ term dominates. The decrease in the
amount of the non-Gaussianity means that the perturbations become
smaller compared to the background value. In our case this would
imply that after inflation the perturbations are damped faster than
the background value. Indeed, we see that this is the case by
considering the first order equation of motion for the
perturbations,
$\delta\ddot{\sigma}+\frac{3}{2t}\delta\dot{\sigma}+V''(\sigma)\delta\sigma=0$,
with a potential correction of the form $\lambda
m_{\sigma}^{4-n}\sigma^{n}$. If the potential is purely quadratic
the perturbations and the background field apparently obey the same
equation.

The increase of the exponent $n$ diminishes the energy density of
the background field at the beginning of oscillations since we are
keeping $s$ fixed. Furthermore, the energy density associated with
the perturbations at the end of inflation gets bigger. With a large
enough $n$ these effects become dominant over the damping of the
perturbations whence the amount of non-Gaussianity again begins to
grow. From Fig. \ref{fnlplot} we see that the bigger the value of
$s$, the smaller is the value of $n$ at which the growth begins;
this is of course quite reasonable.

We should point out here that the increase in $|f_{NL}|$ with a
large $n$ typically happens when the potential correction becomes
significant, $s\sim {2}/{n}$. At this point the perturbative
approach breaks down and the result (Eq. (\ref{Fnl})) can  be at
most in a qualitative agreement with the true behaviour of $f_{NL}$.
The drastic increase in the value of $|f_{NL}|$ seen in Fig.
\ref{fnlplot} is partly due to this effect. In Fig. \ref{fnlplot} we
also display the result of a numerical analysis in which we have,
for simplicity, neglected the small change in the time $t_{\rm osc}$
corresponding to the beginning of oscillations. The values of
$|f_{NL}|$ thus obtained are not significantly different from the
perturbative results which is to be expected since only the region
$s\lesssim{2}/{n}$ is shown in Fig. \ref{fnlplot}. The increase in
$|f_{NL}|$ levels out, however.

When the correction $s$ becomes even larger $s\gtrsim{2}/{n}$ one no
longer can ignore the effect on the beginning of oscillations.
Indeed, the perturbations begin to oscillate way before the
background field and the oscillation may not initially take place in
the quadratic part of potential.  We do not consider such large
corrections in detail here but we make some general remarks on the
behaviour of $f_{NL}$ justifying the use of the perturbative results
in the region $s\sim{2}/{n}$ and giving a qualitative understanding
of the region $s\gtrsim{2}/{n}$ not covered by our perturbative
treatment. Using Eq. (\ref{feom}) we find that the condition for a
local extremum in $f_{NL}(n)$ can approximatively be written as
\baq\label{extremum}\sigma_{\rm osc}\sigma_{\rm
osc}''&=&{\sigma'}_{\rm
osc}^2+\frac{1}{m_{\sigma}^2}\Big(\sigma_{\rm
osc}\partial_n\ddot{\sigma}_{\rm osc}''+\sigma_{\rm
osc}''\partial_n\ddot{\sigma}_{\rm osc}-
2\frac{\sigma_{\rm osc}\sigma_{\rm osc}''}{\sigma_{\rm
osc}'}\partial_n\ddot{\sigma}'\Big).\eaq To obtain this result we
have assumed the potential correction to be of the same order of
magnitude as the quadratic part. In the limit of small corrections
$\lambda\ll1$ the expression in parenthesis in Eq. (\ref{extremum})
vanishes and, since $\sigma''_{\rm osc}<0$ by Eq.
(\ref{curvatonvalue}), we see that  in this region $f_{NL}$ is a
monotonously decreasing function of $n$; this is consistent with our
perturbative  result Eq. (\ref{Fnl}). However, when the correction
becomes larger the terms in Eq. (\ref{extremum}) involving time
derivatives are no longer negligible. Thus, for certain values of
$n$ there exist solutions to Eq (\ref{extremum})  implying that the
growth of the non-linearity parameter $|f_{NL}|$ in the region
$s\gtrsim{2}/{n}$ eventually ceases whereafter $|f_{NL}|$ begins to
oscillate. We do not examine the non-Gaussianity in this region more
closely but we point out that $|f_{NL}|$ might become large enough
to exclude some classes of potentials already with the present WMAP
limits \cite{limits}.


\section{Restrictions on the potential correction}

So far we have considered the potential corrections only from a
technical point of view. There are, however, some physical
motivations for choosing non-quadratic potential. Small corrections
of the type $n=2+\epsilon$ typically represent the effects coming
from one-loop corrections due to light degrees of freedom the
curvaton couples to. As we have seen above, these tend to decrease
the amount of non-Gaussianity. Also, it is conceivable that the
curvaton is self-interacting. The $n=4$ term, in particular, would
be interesting since it implies less fine-tuning to satisfy the
smallness condition of the correction $s\ll{2}/{n}$; for $n=4$ one
would not have to require $\lambda\lll 1$ as for the higher order
cases such as might arise, for example, in models
\cite{enqvist,kenqvist} where the curvaton field is considered to be
one of the flat directions of the Minimally Supersymmetric Standard
Model.
%
%
\begin{figure}[!h]
\hspace{-1.6 cm}
\includegraphics[height=6.5cm, width=18 cm]{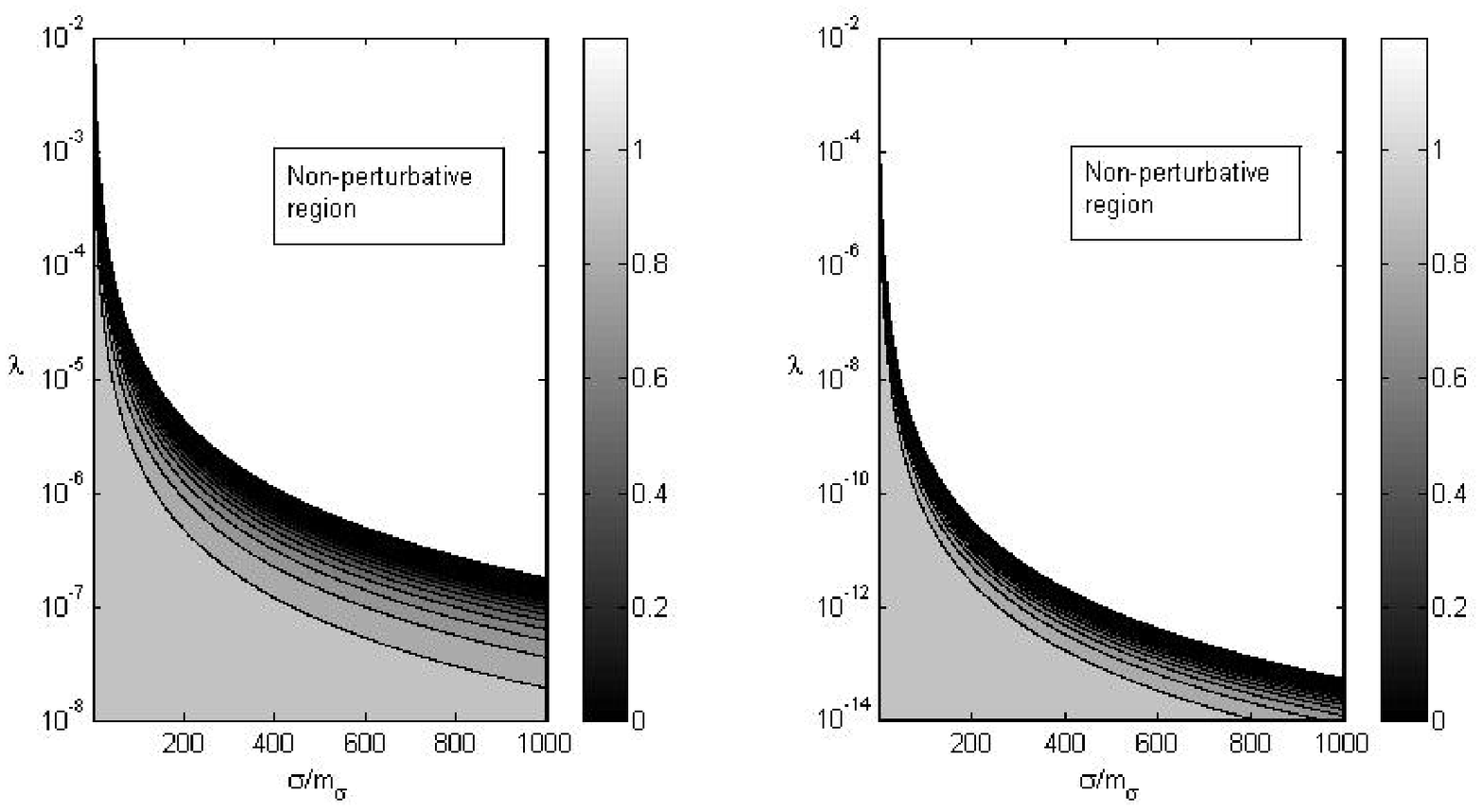}
\caption{\small The non-linearity parameter $|f_{NL}|$ in units
$5/3r=1$ as a contourplot with $n=4$ on the left panel and $n=6$ on
the right panel. The scale on y-axis is logarithmic. The values of
$|f_{NL}|$ are evaluated only in the perturbative region
$s\lesssim2/n$ and the non-perturbative region is printed in white.}
\label{pmspace}
\end{figure}
%
%
%
This is also seen in Fig. \ref{pmspace} where we show $f_{NL}$ as a
contourplot in $(\sigma/m_{\sigma},\lambda)$ space for $n=4$ and
$n=6$. We note that especially in the $n=4$ case there is a
significant region in the parameter space in which the potential
correction is small $s\lesssim2/n$ but the value of $f_{NL}$ is
highly suppressed from the quadratic case. For non-renormalizable
terms $(n>4)$ the requirement of negligible inflaton-generated
curvature perturbations sets an upper limit on $\lambda$ (e.g. for
n=6 $\lambda\lesssim10^{-10}$), but the region in the parameter
space with small values of $|f_{NL}|$ is still considerable. In
other words, it is quite possible to obtain $|f_{NL}|\ll1$ even in
the curvaton models by adding a small self-interaction term to the
quadratic potential.

\begin{figure}[!h]
\begin{center}
\includegraphics[height=7cm, width=7cm]{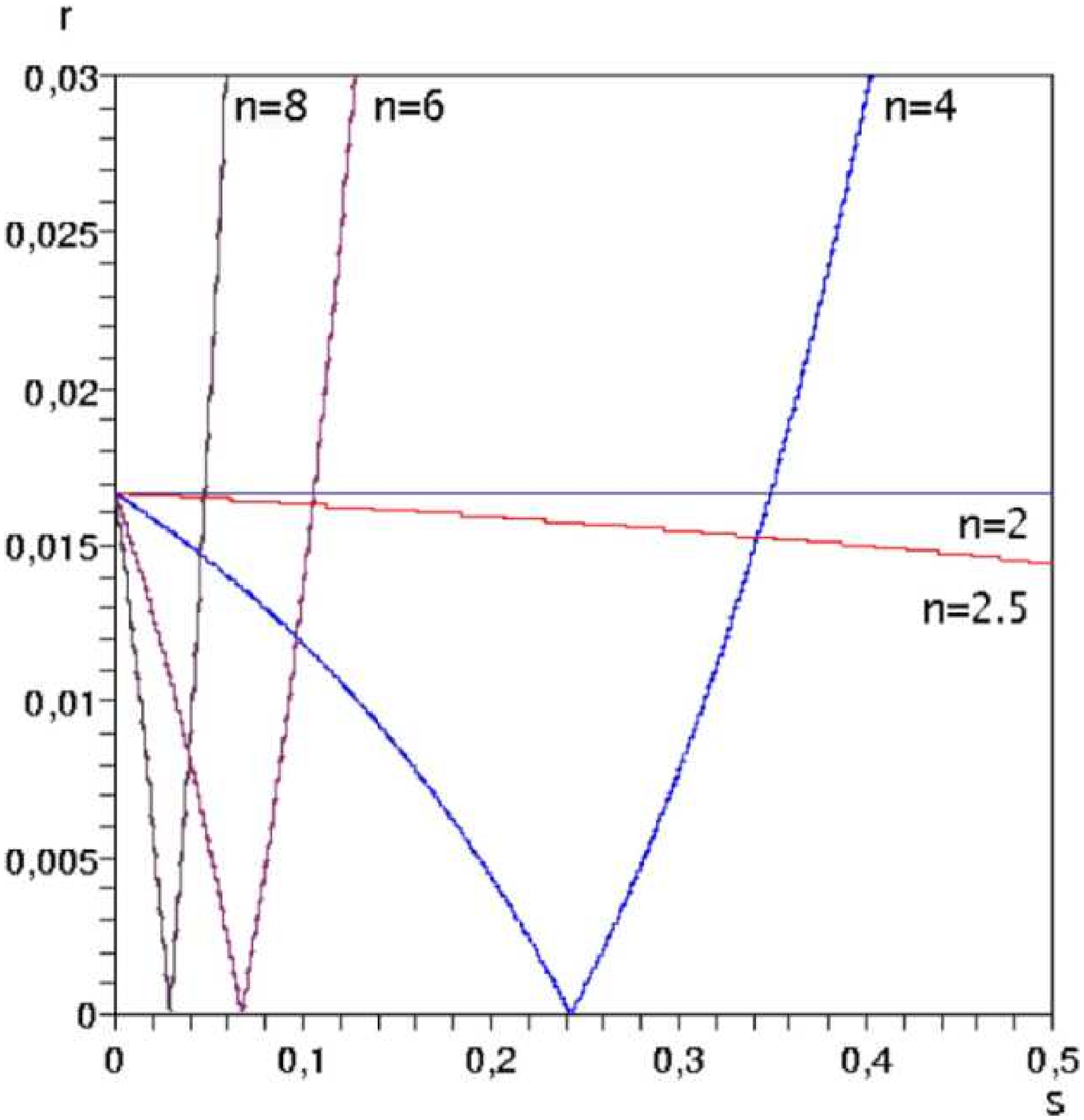}
\caption{\small Allowed regions in the parameter space $(s,r)$. The
labels are outside the allowed region for a given value of $n$.}
\label{obsplot}
\end{center}
\end{figure}

WMAP yields an upper limit \cite{WMAP-nongaussian}
$|f_{NL}|\lesssim100$, which in the case of a quadratic curvaton
potential implies that $r\gtrsim0.02$ \cite{curvatonnonG} as it can
be seen from Eq. (\ref{quadrfnl}). For the non-quadratic case, the
limit is greatly modified due to the decrease in $f_{NL}$. From Eq.
(\ref{Fnl}) the part of the parameter space $(s,r)$ compatible with
present observations is given by \beq
r>\frac{1}{60}\Big|1+\frac{n(n-1)(n-2)g(n)(0.41s+0.25s^2ng(n))}{(0.81+0.50n(n-1)sg(n))^2}\Big|.
\eeq The allowed region is represented in Fig. \ref{obsplot}  for a
choice of parameter values.
%
%
%
%
%
It is noteworthy that the limits on $r$ are strongly dependent on
the size $s$ and form $n$ of the potential correction. In the
non-perturbative region not shown in Fig. \ref{obsplot} we expect
the increase in the lower limit on $r$ to level out as a result of
the behaviour of $f_{NL}$ described above.


\section{Conclusions}

In this paper we have examined how a small departure from the
quadratic curvaton potential will affect the produced level of
non-Gaussianity of the primordial curvature perturbation. For
non-quadratic potentials there are two competing, opposite effects
that contribute to the net non-Gaussianity. First, unlike in the
quadratic case, the perturbations will be damped faster than the
background value. This tends to reduce the value of the
non-linearity parameter $|f_{NL}|$. Second, the increase of the
exponent $n$ diminishes the energy density associated with the
background field at the beginning of oscillations. With for a steep
enough potential this effect becomes dominant and compensates for
the damping of the perturbations. As a consequence, the value of
$|f_{NL}|$ will increase as compared with the purely quadratic case.

The net outcome is that although a small departure from a quadratic
curvaton potential does not modify significantly the Gaussian part
of the perturbation, it can have a pronounced effect on the level of
non-Gaussianity. In particular, the limit $|f_{NL}|\ll1$ is allowed
even with small curvaton energy densities $r\ll1$. This is in sharp
contrast with the quadratic result \cite{curvatonnonG,on} and shows
that the curvaton models are not ruled out by a possible
non-detection of non-Gaussianity as it has been suggested in e.g.
\cite{curvatonnonG}. Furthermore, unlike in the quadratic case,
there are no strict limits that could be placed on the curvaton
energy density parameter $r$. By adding a small correction to the
potential, in practice one can enable arbitrarily small $r$ with
suitably chosen parameter values. This is an interesting result in
the sense that it implies that one can not use present observations
to fix the lower limit for the energy scale of an approximately
quadratic curvaton potential without making further assumptions
about the higher order terms.

\acknowledgments{We thank Yeinzon Rodriguez for helpful discussion
on the requirements implied by the negligible inlaton-generated
curvature perturbation.}

\end{document}